\documentclass[aps,prl,twocolumn,superscriptaddress,nofootinbib]{revtex4}
\usepackage{graphicx,amsmath}

\def\Journal#1#2#3#4{{#1} {\bf #2}, #3 (#4)}

\def\NIMA{{ Nucl. Instrum. Methods} A}
\def\NPA{{ Nucl. Phys.} A}
\def\NPB{{ Nucl. Phys.} B}
\def\NPBproc{{ Nucl. Phys.} B (Proc. Suppl.)}

\def\PLB{{ Phys. Lett.} B}
\def\PRL{ Phys. Rev. Lett.}
\def\PRD{{ Phys. Rev.} D}
\def\ZPC{{ Z. Phys.} C}
\def\EUPJ{{ Eur. Phys. J.} C}

\newcommand{\la}{\langle}
\newcommand{\ra}{\rangle}
\newcommand{\pms} {\hspace{-0.5mm}\pm\hspace{-0.5mm}}
\newcommand{\etal}{{\it et al.}}
\begin{document}

\title{Measurement of single-spin azimuthal asymmetries 
in semi-inclusive electroproduction of pions 
and kaons on a longitudinally polarised deuterium target}


\def\groupalberta{\affiliation{Department of Physics, University of Alberta, Edmonton, Alberta T6G 2J1, Canada}}
\def\groupargonne{\affiliation{Physics Division, Argonne National Laboratory, Argonne, Illinois 60439-4843, USA}}
\def\groupbari{\affiliation{Istituto Nazionale di Fisica Nucleare, Sezione di Bari, 70124 Bari, Italy}}
\def\groupbeijing{\affiliation{Department of Physics, Peking University, Beijing 100871, China}}
\def\groupcolorado{\affiliation{Nuclear Physics Laboratory, University of Colorado, Boulder, Colorado 80309-0446, USA}}
\def\groupdesy{\affiliation{DESY, Deutsches Elektronen-Synchrotron, 22603 Hamburg, Germany}}
\def\groupzeuthen{\affiliation{DESY Zeuthen, 15738 Zeuthen, Germany}}
\def\groupdubna{\affiliation{Joint Institute for Nuclear Research, 141980 Dubna, Russia}}
\def\grouperlangen{\affiliation{Physikalisches Institut, Universit\"at Erlangen-N\"urnberg, 91058 Erlangen, Germany}}
\def\groupferrara{\affiliation{Istituto Nazionale di Fisica Nucleare, Sezione di Ferrara and Dipartimento di Fisica, Universit\`a di Ferrara, 44100 Ferrara, Italy}}
\def\groupfrascati{\affiliation{Istituto Nazionale di Fisica Nucleare, Laboratori Nazionali di Frascati, 00044 Frascati, Italy}}
\def\groupfreiburg{\affiliation{Fakult\"at f\"ur Physik, Universit\"at Freiburg, 79104 Freiburg, Germany}}
\def\groupgent{\affiliation{Department of Subatomic and Radiation Physics, University of Gent, 9000 Gent, Belgium}}
\def\groupgiessen{\affiliation{Physikalisches Institut, Universit\"at Gie{\ss}en, 35392 Gie{\ss}en, Germany}}
\def\groupglasgow{\affiliation{Department of Physics and Astronomy, University of Glasgow, Glasgow G128 QQ, United Kingdom}}
\def\groupillinois{\affiliation{Department of Physics, University of Illinois, Urbana, Illinois 61801, USA}}
\def\groupliverpool{\affiliation{Physics Department, University of Liverpool, Liverpool L69 7ZE, United Kingdom}}
\def\groupwisconsin{\affiliation{Department of Physics, University of Wisconsin-Madison, Madison, Wisconsin 53706, USA}}
\def\groupmit{\affiliation{Laboratory for Nuclear Science, Massachusetts Institute of Technology, Cambridge, Massachusetts 02139, USA}}
\def\groupmichigan{\affiliation{Randall Laboratory of Physics, University of Michigan, Ann Arbor, Michigan 48109-1120, USA }}
\def\groupmoscow{\affiliation{Lebedev Physical Institute, 117924 Moscow, Russia}}
\def\groupmunich{\affiliation{Sektion Physik, Universit\"at M\"unchen, 85748 Garching, Germany}}
\def\groupnikhef{\affiliation{Nationaal Instituut voor Kernfysica en Hoge-Energiefysica (NIKHEF), 1009 DB Amsterdam, The Netherlands}}
\def\groupstpetersburg{\affiliation{Petersburg Nuclear Physics Institute, St. Petersburg, Gatchina, 188350 Russia}}
\def\groupprotvino{\affiliation{Institute for High Energy Physics, Protvino, Moscow region, 142284 Russia}}
\def\groupregensburg{\affiliation{Institut f\"ur Theoretische Physik, Universit\"at Regensburg, 93040 Regensburg, Germany}}
\def\grouprome{\affiliation{Istituto Nazionale di Fisica Nucleare, Sezione Roma 1, Gruppo Sanit\`a and Physics Laboratory, Istituto Superiore di Sanit\`a, 00161 Roma, Italy}}
\def\groupsimonfraser{\affiliation{Department of Physics, Simon Fraser University, Burnaby, British Columbia V5A 1S6, Canada}}
\def\grouptriumf{\affiliation{TRIUMF, Vancouver, British Columbia V6T 2A3, Canada}}
\def\grouptokyo{\affiliation{Department of Physics, Tokyo Institute of Technology, Tokyo 152, Japan}}
\def\groupamsterdam{\affiliation{Department of Physics and Astronomy, Vrije Universiteit, 1081 HV Amsterdam, The Netherlands}}
\def\groupwarsaw{\affiliation{Andrzej Soltan Institute for Nuclear Studies, 00-689 Warsaw, Poland}}
\def\groupyerevan{\affiliation{Yerevan Physics Institute, 375036 Yerevan, Armenia}}


\groupalberta
\groupargonne
\groupbari
\groupbeijing
\groupcolorado
\groupdesy
\groupzeuthen
\groupdubna
\grouperlangen
\groupferrara
\groupfrascati
\groupfreiburg
\groupgent
\groupgiessen
\groupglasgow
\groupillinois
\groupliverpool
\groupwisconsin
\groupmit
\groupmichigan
\groupmoscow
\groupmunich
\groupnikhef
\groupstpetersburg
\groupprotvino
\groupregensburg
\grouprome
\groupsimonfraser
\grouptriumf
\grouptokyo
\groupamsterdam
\groupwarsaw
\groupyerevan


\author{A.~Airapetian}  \groupyerevan
\author{N.~Akopov}  \groupyerevan
\author{Z.~Akopov}  \groupyerevan
\author{M.~Amarian}  \groupzeuthen \groupyerevan
\author{V.V.~Ammosov}  \groupprotvino
\author{A.~Andrus}  \groupillinois
\author{E.C.~Aschenauer}  \groupzeuthen
\author{W.~Augustyniak}  \groupwarsaw
\author{H.~Avakian}  \groupfrascati
\author{R.~Avakian}  \groupyerevan
% ...

\collaboration{The HERMES Collaboration} \noaffiliation

\date{February 18, 2003}

\begin{abstract}

Single-spin asymmetries have been measured for semi-inclusive electroproduction
of $\pi^+$, $\pi^-$, $\pi^0$ and $K^+$ mesons in deep-inelastic scattering off
a longitudinally polarised deuterium target. The asymmetries appear in the
distribution of the hadrons in the azimuthal angle $\phi$ around the virtual
photon direction, relative to the lepton scattering plane. The corresponding
analysing powers in the $\sin \phi$ moment of the cross section are
 $0.012 \pm 0.002 \mbox{(stat.)} \pm 0.002 \mbox{(syst.)}$ for $\pi^+$,
 $0.006 \pm 0.003 \mbox{(stat.)} \pm 0.002 \mbox{(syst.)}$ for $\pi^-$, 
 $0.021 \pm 0.005 \mbox{(stat.)} \pm 0.003 \mbox{(syst.)}$ for $\pi^0$ and 
 $0.013 \pm 0.006 \mbox{(stat.)} \pm 0.003 \mbox{(syst.)}$ for $K^+$.
The $\sin 2\phi$ moments are compatible with zero for all particles. 


\end{abstract} 

 \maketitle


Deep-inelastic lepton scattering (DIS) on polarised nucleons has provided much
of our present understanding of the spin structure of the nucleon.
Recently, measurements of single-spin azimuthal asymmetries have been
recognised as a powerful source of information about the spin structure
of the nucleon~\cite{Mulders+:1996}, complementary to inclusive deep-inelastic
scattering. Significant azimuthal target-spin asymmetries in electroproduction
of $\pi^+$ and $\pi^0$ mesons on a longitudinally polarised hydrogen target
have been reported in Refs.~\cite{hermes_pi+_ssa:1999,hermes_pi0_ssa:2001}.
Evidence for azi\-muthal asymmetries of pions has also been reported for
deep-inelastic lepton scattering off transversely polarised
protons~\cite{smc_ssa:1999}. 

It has been suggested~\cite{Collins:1993} that these single-spin asymmetries
may provide information on the transversity distribution, which describes in a
transverse polarisation basis the probability to find a quark with its spin
parallel or antiparallel to the spin of the nucleon that is polarised
transversely to its (infinite)
momentum~\cite{Ralston-Soper:1979,Artru:1990,Jaffe-Ji:1991}. Transversity is a
chiral-odd distribution function, which implies that it is not observable in an
inclusive measurement, because chirality is conserved in electromagnetic and
strong interactions in the limit of massless on-shell quarks. Therefore, a
second chiral-odd object has to be involved in the
process~\cite{Kotzinian:1995,Mulders:1997}. In semi-inclusive scattering this
can be a chiral-odd fragmentation function --- for example the 
Collins-function~\cite{Collins:1993}. 

The HERMES results on target single-spin
asymmetries~\cite{hermes_pi+_ssa:1999,hermes_pi0_ssa:2001} have elicited a
number of phenomenological studies to evaluate these asymmetries in the
framework of the Collins mechanism using various models as input for the
chiral-odd distribution and fragmentation
functions~\cite{Oganessyan:1998,Boer:1999,Boglione+:2000,Efremov+:2000,deSanctis+:2000,Anselmino+:2000,Karo+:2000,Karo+:2001,Ma+:2001,Efremov+:2001,Ma+:2002}.
Theoretical predictions have also been made for single-spin asymmetries in DIS
off the nucleons in a deuterium target~\cite{Ma+:2002,Efremov+:2002}. 

Recently, it has been shown that another mechanism can also cause a 
single-spin azimuthal asymmetry in semi-inclusive deep-inelastic
scattering~\cite{Brodsky+:2002}. In this case, the observed asymmetry is 
attributed to the interaction of the struck quark with the target remnant
through the exchange of a single gluon. This mechanism was shown to be
identical~\cite{Collins:2002} to the Sivers effect known already for a long
time~\cite{Sivers:1990}, involving a chiral-even time-odd distribution
function. Other theoretical
studies~\cite{Anselmino+:1995,Boer+:1997,Boglione+:1999,Ji+:2002} have revealed
that factorisation applies to this process, which leads to gauge-invariant
momentum dependent parton distributions. In the case of a longitudinally 
polarised target the Collins and the Sivers mechanism cannot be distinguished. 
However, for the two mechanisms a different kinematic dependence on the 
fractional energy $z$ of the hadron has been predicted~\cite{Boglione+:1999}.

This paper reports the first observation of target-spin azimuthal asymmetries 
for semi-inclusive pion and kaon production on a longitudinally polarised
deuterium target. The data were recorded during the 1998, 1999 and 2000 running
periods of the HERMES experiment. The experiment was performed with a beam of
$27.6$~GeV polarised electrons/positrons from the HERA storage ring at DESY and
polarised nucleons in a deuterium gas target. The average target polarisation
was 0.84 with a fractional uncertainty of 5\%. The data were taken with an
electron beam in 1998 and with a positron beam in 1999 and 2000. The measured
single-spin asymmetries show no dependence on the beam charge. Therefore, all
datasets were combined.  In the following, electrons and positrons will be
jointly referred to as positrons.

\begin{figure}[!tb]
\center
\includegraphics[width=\columnwidth]{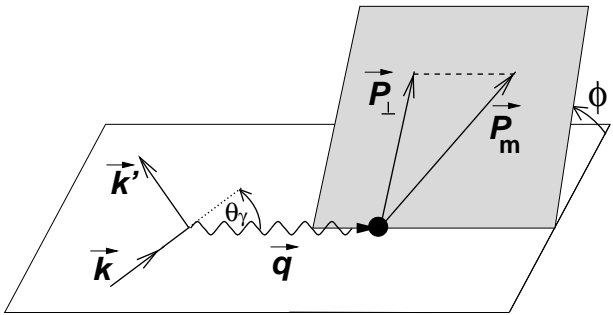}
\caption{Kinematic planes for meson production in semi-inclusive
deep-inelastic scattering: Lepton scattering plane \textit{(white)} and
the meson production plane \textit{(shaded)}.}
\label{figure1}
\end{figure}

The process considered is the production of a pseudoscalar meson ($m=\pi$ or
$K$) in deep-inelastic positron scattering off a longitudinally polarised
deuterium target:
\begin{equation}
e + \overrightarrow{d} \rightarrow e + m + X. 
\end{equation} 
The kinematics of this scattering process are illustrated in 
Fig.~\ref{figure1}. The relevant variables are the squared four-momentum
$-Q^2=q^2=(k-k')^2$ and the energy $\nu=E-E'$ of the virtual photon in the
target rest frame and its fractional energy $y=\nu/E$, the invariant mass of
the virtual-photon nucleon system $W=\sqrt{2M\nu+M^2-Q^2}$, the Bjorken
variable $x=Q^2/2M\nu$ and the fractional energy $z=E_m/\nu$ of the produced
meson. Here, $k$ ($k'$) and $E$ ($E'$) are the 4-momenta and the energies of
the incident (scattered) positron and $M$ is the nucleon mass. The energy and
momentum of the meson in the target rest frame are given by $E_m$ and $P_m$,
respectively. The transverse momentum $P_\perp$ of the produced meson is
defined with respect to the virtual-photon direction. The 
angle $\phi$ is the azimuthal angle of the scattered meson around the virtual
photon direction 
with respect to the lepton scattering plane. Its magnitude is evaluated by  
\begin{equation} 
\cos \phi=\frac{(\overrightarrow{q}\times\overrightarrow{k})
\cdot(\overrightarrow{q}\times
\overrightarrow{P_m})}{|\overrightarrow{q}\times\overrightarrow{k}|\;\;
|\overrightarrow{q}\times \overrightarrow{P_m}|}  
\end{equation}
and its sign by $(\overrightarrow{q}\times \overrightarrow{k}) \cdot
\overrightarrow{P_m}/|(\overrightarrow{q}\times \overrightarrow{k})\cdot
\overrightarrow{P_m}|$. In the case of a target polarised longitudinally with
respect to the incident positron direction, the target polarisation vector has
components parallel and orthogonal with respect to the virtual photon. The
longitudinal and the transverse component of the target polarisation vector are
given by $\cos \theta_\gamma$ and $\sin\theta_\gamma$, respectively. Here,
$\theta_\gamma$ is the angle between the incident positron and the virtual
photon in the photon-nucleon centre-of-mass system. In the HERMES acceptance,
the mean values of $\la \cos \theta_\gamma \ra$ and $\la \sin\theta_\gamma \ra$
are $0.98$ and $0.16$, respectively.

For the measurement of a single target-spin asymmetry the positron beam has to
be unpolarised. The positrons in the HERA storage ring are naturally
transversely polarised by the emission of synchrotron
radiation~\cite{Sokolov+:1964}. The transverse beam polarisation is transformed
into longitudinal polarisation and back to transverse polarisation by two
spin-rotators~\cite{Barber+:1995} upstream and downstream of the HERMES
experiment, respectively. The sign of the beam polarisation is changed about
every two months, which requires moving the magnets of the spin rotators and
inverting their magnetic field direction. The transverse and the longitudinal
positron polarisation are continuously monitored by two Compton-backscattering
polarimeters~\cite{TPOL:1993,LPOL:2002}. To obtain an unpolarised beam, a
polarisation and luminosity weighted average is formed from data of periods
with opposite beam spin orientations. The averaged luminosity weighted beam
polarisation in the analysed data sample is $0.0$\%$ \pm 0.1$\%$ \mbox{(stat.)}
\pm 2.0$\%$ \mbox{(syst.)}$.

The scattered positrons and associated mesons are detected by the HERMES
spectrometer~\cite{hermes:spectrometer} in the range $0.04$~rad~$< \theta <
0.22$~rad of the polar angle. Positron and hadron separation is based on the
information from four detectors: a transition-radiation detector, a 
dual-radiator ring imaging \v{C}erenkov detector (RICH)~\cite{RICH:2002}, a
preshower scintillation detector and a lead-glass electromagnetic
calorimeter~\cite{hermes-calo:1996}. This system provides an average positron
identification efficiency exceeding 98\% with a hadron contamination below 1\%.

Events are required to contain only one electron or positron track with the
same charge as the beam particle and in addition at least one meson. If more
than one meson is detected in the spectrometer, only the meson with the largest
momentum is considered. Identification of charged  pions or kaons in the
momentum range $2$~GeV $<P_m < 15$~GeV is accomplished using the information
from the RICH. Based on a Monte Carlo simulation of the RICH, detection
efficiencies and contaminations for charged pions and kaons are determined 
as a function of the hadron momentum and the hadron multiplicity. The average 
identification efficiency in the RICH is 97\% for pions and 88\% for kaons. 
The detector properties are used to unfold the true hadron populations 
from the measured ones.

\begin{figure}[!tb]   \center 
\includegraphics[width=\columnwidth]{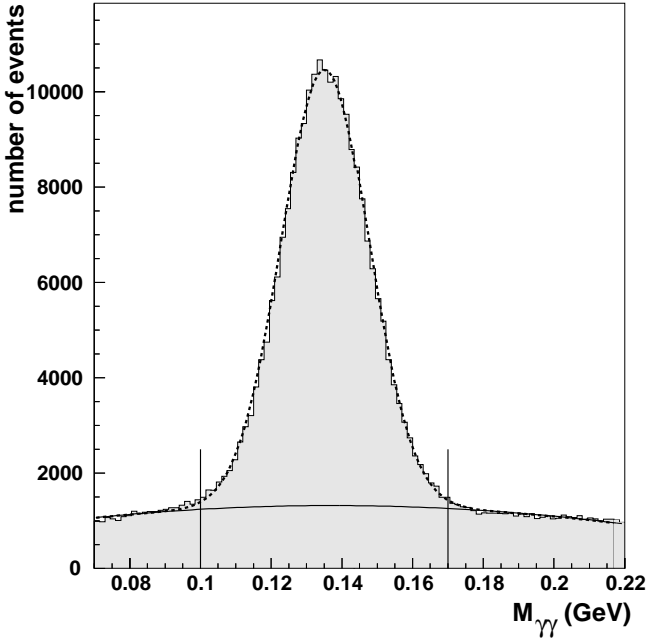}
\caption{Invariant mass ($M_{\gamma\gamma}$) spectrum of photon pairs measured
in the electromagnetic calorimeter. A fit to the data using a Gaussian function
for the peak plus a second order polynomial (solid curve) for the background of
uncorrelated photons is shown as the dotted curve. The two vertical lines
embrace the invariant mass interval used for $\pi^0$ identification.}  
\label{figure2}  
\end{figure} 

Neutral pions are identified by the detection of two photons in the
electromagnetic calorimeter. The reconstructed energy for each photon is 
required to be at least 1.0~GeV and each photon hit in the calorimeter is 
required not to be associated with any charged particle track going in the same
direction. The reconstructed photon-pair invariant mass $M_{\gamma\gamma}$
distribution shows a clear $\pi^0$ mass peak with a mass resolution of about
$0.012$~GeV, as displayed in Fig.~\ref{figure2}. Neutral pions are selected
within the invariant mass range $0.10$~GeV $< M_{\gamma\gamma} < 0.17$~GeV. The
background contribution from uncorrelated photons to the reconstructed
invariant mass spectrum decreases with increasing $z$ of the hadron and ranges 
from 35\% for the lowest $z$ bin to less than 5\% for the highest bin. The
asymmetry of this background is determined outside of the mass window of the
$\pi^0$ mass peak and is found to be compatible with zero. A correction is
applied to account for this dilution.

The requirements imposed on the kinematics of the scattered positron are the
same as those in the previous analyses of the hydrogen
data~\cite{hermes_pi+_ssa:1999,hermes_pi0_ssa:2001}: $1$~GeV$^2 < Q^2 <
15$~GeV$^2$, $W > 2$~GeV, $0.023 < x < 0.4$ and $y < 0.85$. Contributions from
target fragmentation are suppressed by requiring $z> 0.2$ and exclusive meson
production is suppressed by the cut $z < 0.7$. A lower limit of $50$~MeV is
imposed on $P_\perp$ to ensure an accurate measurement of the azimuthal angle
$\phi$.

The target-spin asymmetry $A_\mathrm{UL}$ in the cross section of scattering an
unpolarised beam (U) on a longitudinally polarised target (L) is evaluated as 
\begin{equation}
A_\mathrm{UL}(\phi)=\frac{\displaystyle 1}{|P_\mathrm{L}|}\cdot\frac{
\displaystyle {N^\rightarrow}(\phi)/{L^\rightarrow} -
\displaystyle {N^\leftarrow}(\phi)/{L^\leftarrow}}{
\displaystyle {N^\rightarrow}(\phi)/{L^\rightarrow} +
\displaystyle {N^\leftarrow}(\phi)/{L^\leftarrow}}\,,
\label{equ:AUL}
\end{equation}
where $N^{\rightarrow (\leftarrow)}$ is the number of pions or kaons detected
for target spin antiparallel (parallel) to the direction of the beam momentum,
$L^{\rightarrow (\leftarrow)}$ is the respective dead-time corrected luminosity, 
and $P_\mathrm{L}$ the average longitudinal target polarisation. 
The asymmetry for $\pi^0$ mesons is corrected for the dilution from the
background of uncorrelated photons using the equation 
\begin{equation}
A_\mathrm{corr}(\phi)=
\frac{N_\mathrm{\pi^0}+N_\mathrm{bg}}{N_\mathrm{\pi^0}}\cdot
A_\mathrm{meas}(\phi)  -\frac{N_\mathrm{bg}}{N_\mathrm{\pi^0}}\cdot
A_\mathrm{bg}(\phi)\,.  
\end{equation}  
Here $N_\mathrm{\pi^0}$ and $N_\mathrm{bg}$ are the number of neutral pions and
background-photon pairs, respectively, in each kinematic bin. The asymmetries
for $\pi^0$ mesons $A_\mathrm{meas}$ and for the background of uncorrelated
photons $A_\mathrm{bg}$ are calculated as defined in Eq.~(\ref{equ:AUL}). The
background asymmetry is found to be consistent with zero.

\begin{figure}[!tb]
\center
\includegraphics[width=\columnwidth]{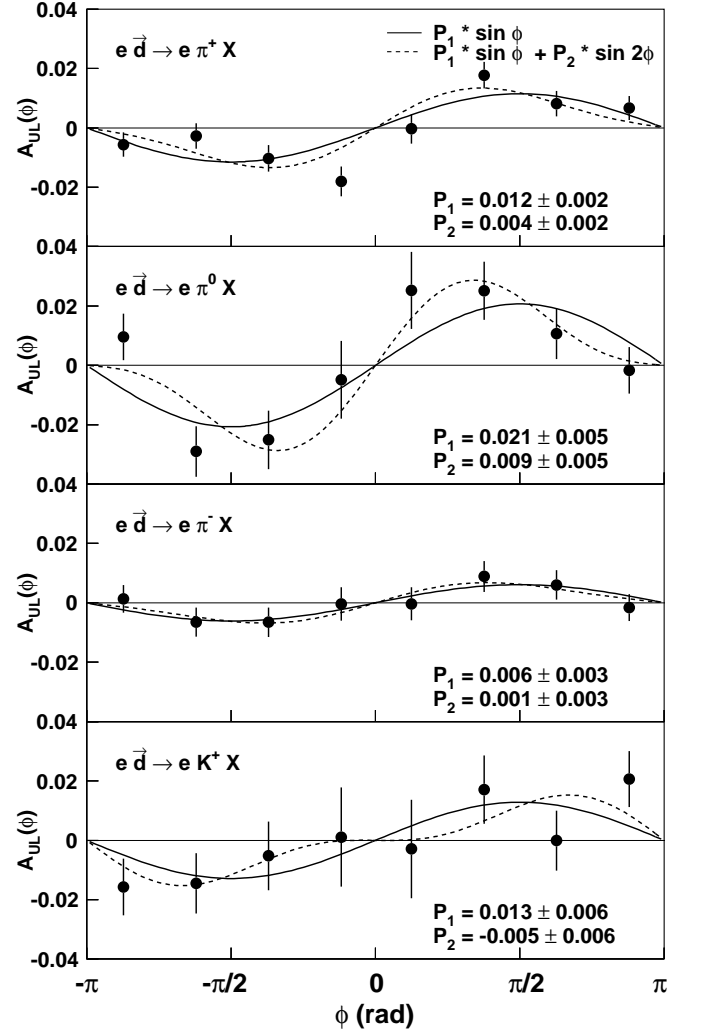} 
\caption{Target spin asymmetries $A_\mathrm{UL}(\phi)$ for electroproduction of
$\pi^+$, $\pi^0$, $\pi^-$ and $K^+$ mesons. Fits of the form $P_0+P_1 \sin\phi$
\textit{(solid line)} and $P_0+P_1 \sin\phi+P_2 \sin2\phi$ \textit{(dashed
line)} are also displayed in the figure.  The error bars give the statistical
uncertainties of the measurements.  The values of the coefficients $P_0$ are
all compatible with zero and the  coefficients $P_1$ and $P_2$ for the various
hadrons and their  statistical uncertainties are listed in each panel.}
\label{figure3}
\end{figure}

In Fig.~\ref{figure3}, the azimuthal asymmetries $A_\mathrm{UL}(\phi)$ for the
mesons $\pi^+$, $\pi^0$, $\pi^-$ and $K^+$ are displayed as a function of
$\phi$, integrated over the experimental acceptance in the kinematic variables
$x$, $P_\perp$, $z$, $y$ and $Q^2$. The average values are  $\la x\ra=0.09$,
$\la P_\perp \ra=0.40$~GeV, $\la z \ra=0.38$, $\la y\ra =0.53$ and $\la
Q^2\ra=2.4$ GeV$^2$. 

The asymmetries defined in Eq.~(\ref{equ:AUL}) were alternatively fit with 
the functions 
\begin{align}
f_1(\phi)&=P_0+P_1 \sin\phi \hspace{2cm}\\
f_2(\phi)&=P_0+P_1 \sin\phi+P_2\sin2\phi \; , \label{equ:fit2}
\end{align}
which are indicated as curves in Fig.~\ref{figure3}. All
coefficients $P_0$ are compatible with zero. The $\sin\phi$ and $\sin2\phi$
amplitudes $P_1$ and $P_2$, obtained from the fit~(\ref{equ:fit2}) to the 
data, are displayed in the figure as well. They represent the analysing 
powers $A_\mathrm{UL}^{\sin
\phi}$ and $A_\mathrm{UL}^{\sin 2 \phi}$ of the azi\-muthal asymmetry. The
numerical values are given in Tab.~\ref{table1} for the
various mesons, together with the previously reported analysing powers for pion
production on longitudinally polarised protons~\cite{hermes_pi+_ssa:1999,hermes_pi0_ssa:2001}. 

\begin{table}[!tb]
\center
\begin{tabular}{c|c|c|c}
\hline \hline 
 & meson & deuterium target & proton target 
\cite{hermes_pi+_ssa:1999,hermes_pi0_ssa:2001} \\
\hline
$A_{\mathrm{UL}}^{\sin\phi}$ & $\pi^+$ & $\phantom{-}0.012 \pms 0.002 \pms 0.002$ & 
     $\phantom{-}0.022 \pms 0.005 \pms 0.003$ \\
& $\pi^0$ & $\phantom{-}0.021 \pms 0.005 \pms 0.003$ &
     $\phantom{-}0.019 \pms 0.007 \pms 0.003$ \\
& $\pi^-$ & $\phantom{-}0.006 \pms 0.003 \pms 0.002$ & 
     $-0.002\pms 0.006 \pms 0.004$ \\
& $K^+$ & $\phantom{-}0.013 \pms 0.006 \pms 0.003$ & --- \\
\hline 
$A_{\mathrm{UL}}^{\sin2\phi}$ & $\pi^+$ & $\phantom{-}0.004 \pms 0.002 \pms 0.002$ & 
     $-0.002 \pms 0.005 \pms 0.003$ \\
& $\pi^0$ &$\phantom{-}0.009 \pms 0.005 \pms 0.003$ &
     $\phantom{-}0.006 \pms 0.007 \pms 0.003$ \\
& $\pi^-$ & $\phantom{-}0.001 \pms 0.003 \pms 0.002$ & 
     $-0.005\pms 0.006 \pms 0.005$ \\
& $K^+$ & $-0.005 \pms 0.006 \pms 0.003$ & --- \\
\hline \hline
\end{tabular}
\caption{Analysing powers $A_\mathrm{UL}^\mathrm{\sin\phi}$ and
$A_\mathrm{UL}^{\sin 2\phi}$ for the azi\-muthal target-spin asymmetry for the
electroproduction of pions and kaons on the deuteron, integrated over the
experimental acceptance in $x$, $P_\perp$,  $z$, $y$ and $Q^2$. Also listed are
earlier  results obtained on the proton from
Ref.~\cite{hermes_pi+_ssa:1999,hermes_pi0_ssa:2001}. The first uncertainty is
the statistical and the second is the systematic uncertainty of the
measurement.}
\label{table1}
\end{table}

The effects of smearing and spectrometer acceptance are
estimated using a Monte Carlo simulation. For this purpose,
a Monte Carlo simulation is carried out with various $x$,
$P_\perp$ or $z$ dependent $\sin \phi$ and $\sin 2\phi$
amplitudes. Within the statistical accuracy, the
reconstructed event distributions show the same $\sin\phi$
and $\sin 2\phi$ amplitudes as the generated distributions.
It is concluded from the Monte Carlo simulation that the
measured asymmetries are not affected by acceptance or
smearing effects of the detector in any significant way
within the statistical precision of the Monte Carlo
simulation of $0.001$ $(0.002)$ for charged mesons
($\pi^0$). 

An additional test of possible acceptance effects was performed using
measurements with unpolarised hydrogen and deuterium gas targets. These 
measurements were regularly done after a few hours of data taking with
polarised targets. The data were analysed with the kinematic requirements
described above and the $\sin\phi$ and $\sin 2\phi$ moments
$A_\mathrm{UU}^{\sin\phi}$ and $A_\mathrm{UU}^{\sin 2\phi}$ of the unpolarised
cross section are extracted. They were calculated respectively as
$A_\mathrm{UU}^{\sin \phi}=1/N\sum_{i=1}^N\sin  \phi_i $ and 
$A_\mathrm{UU}^{\sin 2\phi}=1/N\sum_{i=1}^N\sin  2\phi_i $, summed over all $N$
events taken with unpolarised target gas. The moments 
$A_\mathrm{UU}^{\sin \phi}$ and $A_\mathrm{UU}^{\sin 2\phi}$ were found
to be consistent with zero as expected~\cite{Mulders+:1996} for pions 
(kaons) within a statistical uncertainty of 0.002 (0.004).

The analysing powers $A_\mathrm{UL}^\mathrm{\sin\phi}$ extracted from a fit to
the asymmetry $A_\mathrm{UL}(\phi)$ have been compared to those
obtained as moments:
\begin{equation}
A_\mathrm{UL}^{W}=\frac{1}{|P_\mathrm{L}|}
\frac{\displaystyle \frac{1}{L^\rightarrow}
\sum\limits_{i=1}^{N^\rightarrow}W(\phi_i) -
\frac{1}{L^\leftarrow}\sum\limits_{i=1}^{N^\leftarrow}W(\phi_i) }
{\frac{1}{2}[N^\rightarrow/L^\rightarrow + N^\leftarrow/L^\leftarrow]},
\label{equ:weighting}
\end{equation}
using the weighting functions $W(\phi)=\sin\phi$ and $W(\phi)=\sin2\phi$,
respectively. This type of analysis is more sensitive to the experimental
acceptance~\cite{hermes_pi0_ssa:2001}. Based on a Monte Carlo simulation,
corrections of about 15\% had to be applied to account for a
cross-contamination between the $\sin\phi$ and $\sin2\phi$ moments. After these
corrections, the analysing powers extracted as moments according to
Eq.~(\ref{equ:weighting}) and those extracted using a fit to the cross section
asymmetry $A_\mathrm{UL}(\phi)$ agree within the systematic uncertainty
assigned to effects of the spectrometer acceptance (see
Tab.~\ref{table3}).

\begin{figure*}[p]  \center
\includegraphics[width=0.9\textwidth]{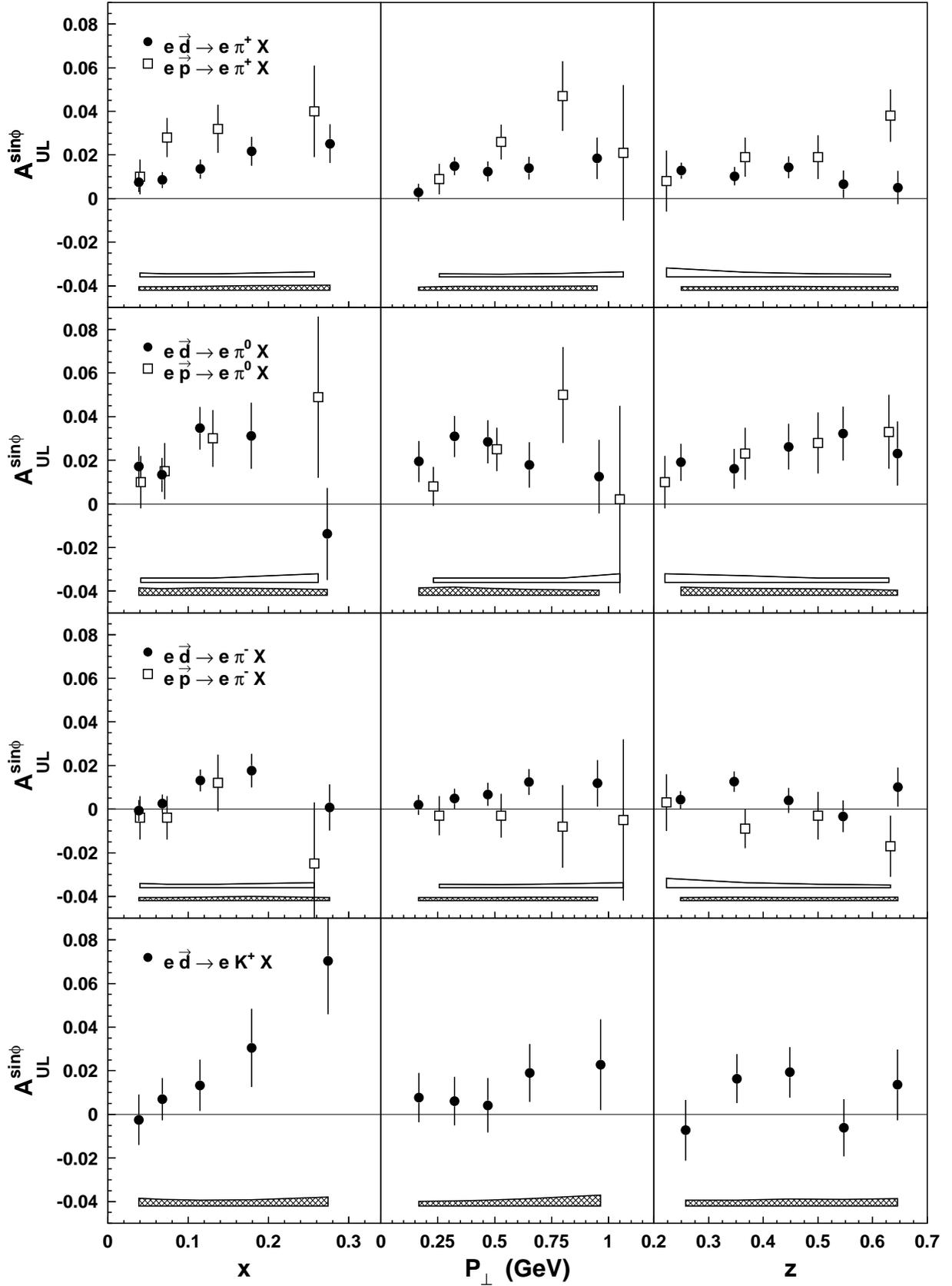} 
\vspace{-0.4cm}
\caption{Target spin analysing powers $A_\mathrm{UL}^\mathrm{\sin\phi}$ for
semi-inclusive $\pi^+$, $\pi^0$, $\pi^-$ and $K^+$ production on the  deuteron 
\textit{(filled circles)} and on the proton \textit{(open squares)}.  The
latter are taken from Refs.~\cite{hermes_pi+_ssa:1999,hermes_pi0_ssa:2001}. The
data are shown as a function of one of the kinematic variables $x$, $P_\perp$
and $z$ while integrating over the other variables. The error bars give the
statistical uncertainties of the measurements and the bands in the lower parts
of each panel give the systematic uncertainties for the deuteron 
\textit{(hashed band)} and for the proton measurement \textit{(open band)}.}
\label{figure4}  
\end{figure*}

In Fig.~\ref{figure4}, the analysing powers
$A_{\mathrm{UL}}^{\sin\phi}$ on the deuteron are shown as a function of $x$,
$P_\perp$ and $z$ together with earlier results obtained on the
proton~\cite{hermes_pi+_ssa:1999,hermes_pi0_ssa:2001}. The mean values of $Q^2$
for each $x$ bin and the mean values of $P_\perp$ for each $z$ bin are given
in Tab.~\ref{table2}. 

\begin{table}[!tb]
\center
\begin{tabular}{l| l l l l l}
\hline \hline
 ~$x$ & ~$0.039$ & ~$0.068$ & ~$0.115$ & ~$0.179$ & ~$0.276$ \\
\hline 
~$\la Q^2\ra$ in GeV$^2$ & ~$1.30$ & ~$1.82$ & ~$2.62$ & ~$3.58$ & ~$4.88$ \\
\hline \hline
\end{tabular}\\
\vspace*{5mm}
\begin{tabular}{l| l l l l l}
\hline \hline
 ~$z$ & ~$0.25$ & ~$0.35$~ & ~$0.45$~ & ~$0.55$~ & ~$0.65$~ \\
\hline 
~$\la P_\perp\ra$ in GeV & ~$0.36$~ & ~$0.40$~ & ~$0.44$~ & ~$0.46$~ & ~$0.47$~ \\
\hline \hline
\end{tabular}
\caption{Mean values of $Q^2$ for each $x$ bin \textit{(upper table)} and mean
values of $P_\perp$ for each bin of $z$ \textit{(lower table)}.}
\label{table2}
\end{table}

\begin{table}[!tb]
\center
\begin{tabular}{l|c|c|c}
\hline \hline 
source of systematic uncertainty & $\pi^+$, $\pi^-$ & $\pi^0$ & $K^+$ \\
\hline
determination of target polarisation & $0.001$ & $0.001$ & $0.001$  \\
upper limit on acceptance effects & $0.001$ & $0.002$ & $0.001$ \\
meson identification (RICH) & $0.0004$ & - & $0.002$ \\
$\rho^0$ contamination  & $0.001$ & - & - \\
$\gamma\gamma$-background correction & - & $0.002$ & - \\
\hline
quadratic sum & $0.002$ & $ 0.003$ & $ 0.003$ \\
\hline \hline
\end{tabular}
\caption{Contributions to the systematic uncertainty of the experimental results
for the target spin analysing powers $A_\mathrm{UL}^{\sin\phi}$ listed in
Tab.~\ref{table1} for $\pi^+$,
$\pi^-$, $\pi^0$ and $K^+$ mesons. The total systematic uncertainty is
calculated as the quadratic sum of the individual contributions.}
\label{table3}
\end{table}

The various contributions to the systematic uncertainty of the experimental
results in Tab.~\ref{table1}, integrated over $x$, $P_\perp$ and $z$, are
listed in Tab.~\ref{table3}. For charged pions, the largest contributions
originate from the determination of the target polarisation and from the upper
limit for possible acceptance effects evaluated in a Monte Carlo simulation.
For kaons the uncertainty in the hadron identification with the RICH detector
also contributes significantly. For pions the RICH efficiency is larger and the
contamination by other hadrons is smaller so that the contribution to the
systematic uncertainty is small. The charged pion sample can be contaminated by
pions from the decay of heavier mesons. The main contribution originates from
the decay of exclusively produced $\rho^0$ vector mesons and is estimated using
a Monte Carlo simulation. It is found to be smaller than 5\%. In addition, it
is shown from the experimental data that there is no asymmetry in their
azimuthal distribution. A contribution is added to the systematic uncertainty
for this dilution. For $\pi^0$ mesons there is a significant contribution to
the systematic uncertainty due to the uncertainty in the determination of the
background yield and its asymmetry. 

The analysing power $A_\mathrm{UL}^{\sin\phi}$ for $\pi^+$ production on the
deuteron is greater than zero, but smaller than that obtained on the proton
(see Tab.~\ref{table1}). In the context of models based on transversity, the
different size of the asymmetries for $\pi^+$ production on the proton and
deuteron can be attributed to the dominant role of the $u$-quark contribution
to the observed asymmetry~\cite{Efremov+:2002}. The analysing powers for
$\pi^0$ production are positive for both deuteron and proton and of similar
size. For $\pi^-$ production, only the deuteron data suggest an asymmetry
different from zero. The result for $K^+$ production on the deuteron  is
compatible with that of $\pi^+$ production, which may indicate the dominant
contribution from $u$-quarks fragmenting into kaons. 

The results for the two targets show a similar behaviour in their kinematic
dependences on $x$, $P_\perp$ and $z$. The observed increase of
$A_\mathrm{UL}^\mathrm{\sin\phi}$ with increasing $x$ suggests that the
single-spin asymmetries are associated with valence quark contributions. 

Two mechanisms have been proposed to explain the measured single-spin
asymmetries. One is the combination of transversity-related chiral-odd 
distribution functions and chiral-odd fragmentation functions like the Collins
fragmentation function. The other one is a final-state interaction of the
struck quark with the target remnant (Sivers
Effect)~\cite{Brodsky+:2002,Ji+:2002}. There are no calculations for a
deuterium target available for the latter scenario that can be compared with
the present data. 

\begin{figure}[p] 
\center 
\includegraphics[width=\columnwidth]{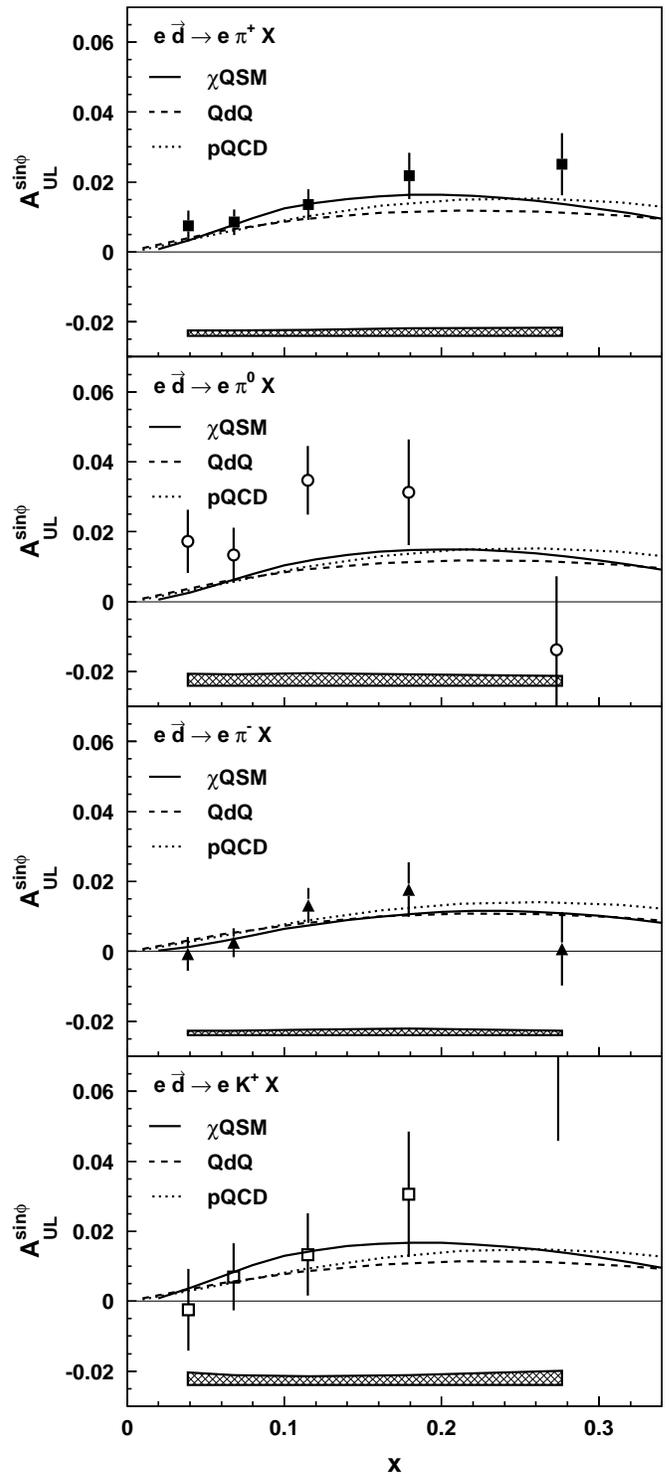} 
\caption{Comparison of the measured analysing powers
$A_\mathrm{UL}^\mathrm{\sin\phi}$ on the deuteron for $\pi^+$, $\pi^0$, 
$\pi^-$ and $K^+$ production with predictions from theoretical  calculations in
the chiral quark soliton model ($\chi$QSM, \textit{solid
lines}~\cite{Efremov+:2002}), the quark-diquark model (QdQ, \textit{dashed
lines}~\cite{Ma+:2002}) and a perturbative QCD model (pQCD, \textit{dotted
lines}~\cite{Ma+:2002}. The shown curves refer to ``approach 2" of the models
in Ref.~\cite{Ma+:2002}. The error bars give the statistical uncertainties of
the measurements, and the bands in the lower part of the panels show the 
systematic uncertainties of the measurements.} 
\label{figure5} 
\end{figure}

Recent model calculations in the context of
transversity~\cite{Ma+:2002,Efremov+:2002} predict $A_\mathrm{UL}^{\sin\phi}$
for scattering on the deuteron within the kinematic range of the HERMES
experiment. These calculations are performed in the same framework than those
mentioned in our earlier publications of the proton
results~\cite{hermes_pi+_ssa:1999,hermes_pi0_ssa:2001}, but take into account a
recently detected sign error in the earlier theoretical
calculations~\cite{Efremov+:2001,Ma+:2001}. The transversity distributions
calculated in the chiral quark soliton model ($\chi$QSM)~\cite{Efremov+:2002},
in the SU(6) quark spectator diquark model~\cite{Ma+:2002} and in a
perturbative QCD model~\cite{Ma+:2002} have been used as an input. The results
of three of these calculations are displayed in Fig.~\ref{figure5} together
with the experimental data. As can be seen from Fig.~\ref{figure5}, the
experimental data are well described by these calculations.

\begin{figure}[!tb]
\center
\includegraphics[width=\columnwidth]{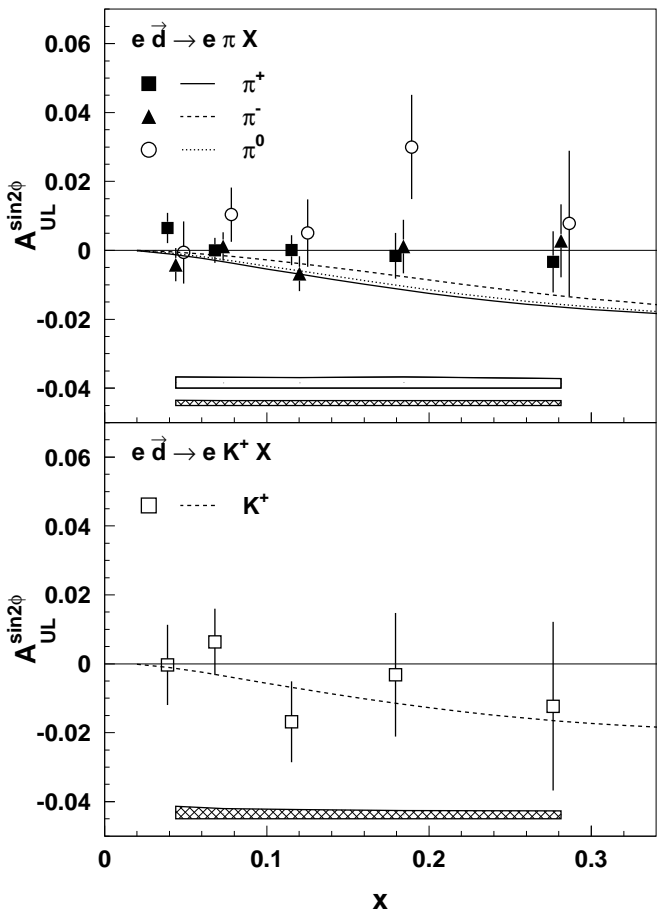}
\caption{The $\sin2\phi$ analysing powers $A_\mathrm{UL}^{\sin2\phi}$ for
$\pi^+$, $\pi^0$ and $\pi^-$ \textit{(upper panel)} and for $K^+$ production \textit{(lower
panel)} on the deuteron. The error bars give the
statistical uncertainties of the measurements. The systematic uncertainties for
$\pi^+$ and $\pi^-$ are represented by the hatched band and those for $\pi^0$ by the
open band. The points for $\pi^0$ and $\pi^-$ are slightly shifted in $x$ for
better visibility. Included as curves are predictions from a
transversity-related calculation in the chiral quark soliton model~\cite{Efremov+:2002}.}
\label{figure6}
\end{figure}

The analysing power $A_{\mathrm{UL}}^{\sin2\phi}$ is an additional
important observable, since it appears as a leading term in the expansion 
of the cross section for scattering electrons off a longitudinally polarised 
target, while the $\sin \phi$ moment appears only at order 
$1/Q$~\cite{Mulders+:1996}. The dependence of
$A_{\mathrm{UL}}^{\sin2\phi}$ on $x$ is presented in Fig.~\ref{figure6}.
Integrated over the measured $x$-range, it is compatible with zero for all
mesons (see Tab.~\ref{table1}). Also shown are corresponding values calculated 
in the $\chi$QSM~\cite{Efremov+:2002}. For pions, the data do not favour the 
predicted trend towards negative asymmetries at large $x$.

\begin{figure}[!tb] \center
\includegraphics[width=\columnwidth]{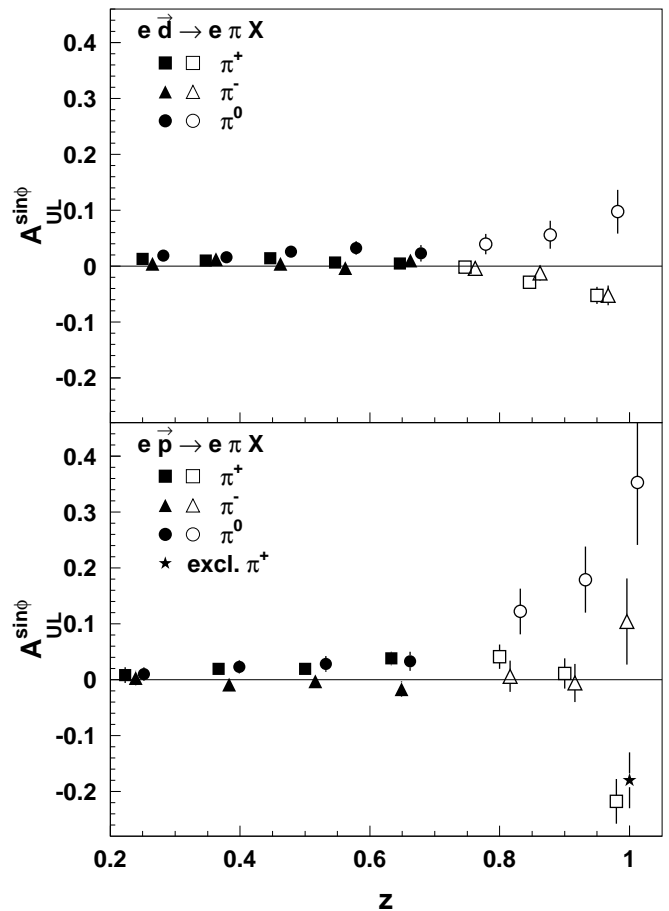}
\caption{The dependence on $z$ of the analysing powers
$A_\mathrm{UL}^\mathrm{\sin\phi}(z)$ for $\pi^+$, $\pi^0$ and $\pi^-$
production on the deuteron \textit{(upper panel)}  and on the proton
\textit{(lower panel)}. The filled symbols show the semi-inclusive measurements
on the deuteron from Fig.~\ref{figure4} and on the proton  from
Ref.~\cite{hermes_pi+_ssa:1999}, respectively. Shown as filled star is the
exclusive measurement for $\pi^+$ production from
Ref.~\cite{hermes_pi+_exclusive:2002}. For the data at high $z$ \textit{(open
symbols)}, no corrections for the experimental resolution in $z$ or possible
contaminations by pions from the decay of exclusive $\rho^0$ vector mesons have
been applied.  The error bars indicate the statistical uncertainty of the
measurements. The points for $\pi^0$ and $\pi^-$ are slightly shifted in $z$
for visibility.} 
\label{figure7}  
\end{figure}

The data presented so far are evaluated in the semi-inclusive kinematic range
$0.2 < z < 0.7$. In Fig.~\ref{figure7}, the $z$-dependencies of the single
spin asymmetries $A_\mathrm{UL}^\mathrm{\sin\phi}$ on the proton and on the
deuteron are shown up to $z=1$. The results on the proton have been obtained
from ex\-perimental data taken with a longitudinally polarised hydrogen target as
described in Ref.~\cite{hermes_pi+_ssa:1999}, neglecting the upper $z<0.7$ cut,
however. The mean experimental resolu\-tion in $z$ is $\Delta z = 0.02$ $(0.04)$
for charged (neutral) pions in the semi-inclusive regime and $\Delta z = 0.07$
$(0.06)$ for $z\rightarrow 1$. It has to be pointed out that the experimental
data shown as open symbols in Fig.~\ref{figure7} have not been corrected for
this variation in $\Delta z$. Also, the results for charged pions have not been
corrected for a possible contamination by pions from the decay of exclusively
produced $\rho^0$ vector mesons.

At large $z$, a transition from the semi-inclusive regime to the exclusive
regime is observed. In the exclusive limit ($z\rightarrow 1$), the scattering
process can be interpreted in terms of generalised parton
distributions~\cite{Mueller+:1994,Radyushkin:1997,Frankfurt+:2000,Belitzky+:2001}.
The data show an inversion of the sign and an increase in absolute size of the
single spin asymmetries, similar for both  $\pi^-$ and $\pi^+$. The size of the
asymmetry for $\pi^0$ mesons increases but it remains positive for all $z$. A
large asymmetry in the exclusive limit has already been reported for exclusive
$\pi^+$ production on the proton~\cite{hermes_pi+_exclusive:2002}. As shown in
the lower panel of Fig.~\ref{figure7}, there is a large analysing power for
$\pi^0$ production on the proton as well, while no significant asymmetry for
$\pi^-$ production is found. No theoretical explanation yet exists for this
experimental result.


In summary, single-spin azimuthal asymmetries for electroproduction of $\pi^+$,
$\pi^0$, $\pi^-$ and $K^+$ mesons on a longitudinally polarised deuterium
target have been measured for the first time. The dependences of these
asymmetries on $x$, $P_\perp$ and $z$ have been investigated. The results show
positive asymmetries for $\pi^+$ and $\pi^0$ and an indication of a positive
asymmetry for $\pi^-$ mesons. The asymmetry for $K^+$ is compatible with that
for $\pi^+$ mesons. These findings can be well described by model calculations
where the asymmetries are interpreted in the context of transversity as the
effect of combinations of chiral-odd distribution functions and chiral-odd
fragmentation functions. Here, the observed asymmetries for $\pi^+$ and $K^+$
are consistent with the assumption of $u$-quark dominance in the quark
distribution and the fragmentation process. Together with earlier measurements
on the proton~\cite{hermes_pi+_ssa:1999,hermes_pi0_ssa:2001}, the results
support the existence of non-zero chiral-odd distribution functions that
describe the transverse polarisation of quarks. However, it cannot be excluded
that a part of the observed asymmetry is due to an additional exchange of a
gluon in the final state (Sivers effect) as discussed in
Ref.~\cite{Brodsky+:2002}. Furthermore, the data show an increase
of the magnitude of the asymmetries for charged and neutral pions at large $z$
when approaching the exclusive regime.

We thank M. Anselmino,  A. Bacchetta, A.M. Kotzinian, P.J. Mulders and P.
Schweitzer for many interesting discussions on this subject. We gratefully
acknowledge the DESY management for its support, the staffs at DESY and the
collabo\-rating institutions for their significant effort. This work was
supported by the FWO-Flanders, Belgium; the Natural Sciences and Engineering
Research Council of Canada; the ESOP, INTAS and TMR network contributions from
the European Union; the German Bundesministerium f\"ur Bildung und
Forschung; the Italian Instituto Nazionale di Fisica Nucleare (INFN); Monbusho
International Scientific Research Program, JSPS and Toray Science Foundation
of Japan; the Dutch Foundation for Fundamenteel Onderzoek der Materie (FOM);
the U.K. Particle Physics and Astronomy Research Council; and the U.S.
Department of Energy and National Science Foundation.

\enlargethispage{1cm}

\end{document}